\documentclass[longauth,letter]{aa}

\usepackage{txfonts}
\usepackage{graphicx}
\usepackage{natbib}
\bibpunct{(}{)}{;}{a}{}{,}

\begin{document}

\title{Discovery of VHE $\gamma$-rays from the distant BL Lac 1ES\,0347-121}

\author{F. Aharonian\inst{1,13}
 \and A.G.~Akhperjanian \inst{2}
 \and U.~Barres de Almeida \inst{8} \thanks{supported by CAPES Foundation, Ministry of Education of Brazil}
 \and A.R.~Bazer-Bachi \inst{3}
 \and B.~Behera \inst{14}
 \and M.~Beilicke \inst{4}
 \and W.~Benbow \inst{1}
 \and K.~Bernl\"ohr \inst{1,5}
 \and C.~Boisson \inst{6}
 \and O.~Bolz \inst{1}
% ...
}

\institute{
Max-Planck-Institut f\"ur Kernphysik,
Heidelberg, Germany
\and
 Yerevan Physics Institute, Yerevan,
Armenia
\and
Centre d'Etude Spatiale des Rayonnements, CNRS/UPS, Toulouse, France
\and
Universit\"at Hamburg, Institut f\"ur Experimentalphysik, Hamburg, Germany
\and
Institut f\"ur Physik, Humboldt-Universit\"at zu Berlin, Berlin, Germany
\and
LUTH, Observatoire de Paris, CNRS, Universit\'e Paris Diderot, Meudon, 
France
\and
DAPNIA/DSM/CEA, CE Saclay, Cedex, France
\and
University of Durham, Department of Physics, Durham,
U.K.
\and
Unit for Space Physics, North-West University, Potchefstroom,
    South Africa
\and
Laboratoire Leprince-Ringuet, Ecole Polytechnique, CNRS/IN2P3,
Palaiseau, France
\and 
Laboratoire d'Annecy-le-Vieux de Physique des Particules, CNRS/IN2P3,
Annecy-le-Vieux, France
\and
Astroparticule et Cosmologie (APC), CNRS, Universite Paris 7 Denis Diderot,
10, Paris, France
\thanks{UMR 7164 (CNRS, Universit\'e Paris VII, CEA, Observatoire de Paris)}
\and
Dublin Institute for Advanced Studies, Dublin,
Ireland
\and
Landessternwarte, Universit\"at Heidelberg, Heidelberg, Germany
% ...
}

\offprints{M. Raue, \email{martin.raue@desy.de}}

\date{Received  / Accepted }

\abstract {}
{Our aim is to study the production mechanism for very-high-energy (VHE; $>$100\,GeV) $\gamma$-rays in distant active galactic nuclei (AGN) and use the observed VHE spectrum to derive limits on the Extragalactic Background Light (EBL). We also want to determine physical quantities through the modeling of  the object's broad-band spectral energy distribution (SED).} 
{VHE observations  ($\sim$25\,h live time) of the BL Lac 1ES\,0347-121 (redshift $z = 0.188$) were conducted with the High Energy Stereoscopic System (H.E.S.S.) between August and December 2006. Contemporaneous X-ray and UV/optical observations from the SWIFT satellite are used to interpret the SED of the source in terms of a synchrotron self Compton (SSC) model.} 
{An excess of 327 events, corresponding to a statistical significance of 10.1 standard deviations, is detected from 1ES\,0347-121. Its photon spectrum, ranging from $\sim$250\,GeV to $\sim$3\,TeV, is well described by a power law with a photon index of $\Gamma = 3.10 \pm 0.23_{\mathrm{stat}} \pm 0.10_{\mathrm{sys}}$. The integral flux above 250\,GeV corresponds to $\sim$2\% of the flux of the Crab Nebula above the same threshold. No VHE flux variability is detected within the data set.}
{Constraints on the EBL density at optical to near-infrared wavelengths derived from the photon spectrum of 1ES\,0347-121 are close to the strongest limits derived previously. The strong EBL limits confirm earlier findings, that the EBL density in the near-infrared is close to the lower limits from source counts. This implies that the universe is more transparent to VHE $\gamma$-rays than previously believed. An SSC model provides a reasonable description of the contemporaneous SED.}


\authorrunning{Aharonian, F.~A. (H.E.S.S. Collaboration)}
\titlerunning{Discovery of VHE $\gamma$- rays from the BL Lac 1ES\,0347-121}

\maketitle

\section{Introduction}

1ES\,0347-121 was discovered in the Einstein Slew Survey \citep{elvis:1992a} and later classified as a BL Lac object \citep{schachter:1993a}. Located at a redshift of $z = 0.188$ (\citealt{woo:2005a}) it harbors a super-massive black hole of mass $\log(M_{\mathrm{BH}}/M_{\mathrm{\odot}}) = 8.02 \pm 0.11$ \citep{woo:2005a}. The host is classified as an elliptical galaxy with luminosity $M_R = -23.2$ \citep{falomo:1999a}. \citet{stecker:1996a} used simple physical considerations about the synchrotron and inverse-Compton component of blazar spectra to calculate a prediction for the VHE $\gamma$-ray flux above 0.3\,TeV of $3.8 \times 10^{-12}$\,cm$^{-2}$\,s$^{-1}$ (0.03 Crab), which would be easily detectable with the current generation of VHE instruments. An upper limit on the integral flux above an energy threshold of 1.46\,TeV of $5.14 \times 10^{-12}$\,cm$^{-2}$\,s$^{-1}$ (0.56 Crab) has been reported by the HEGRA collaboration \citep{aharonian:2004:hegra:agn:ul}, considerably higher than the above mentioned prediction.

For such a distant object, the observed VHE spectrum is expected to be strongly affected by absorption of the $\gamma$-rays due to the EBL \citep{gould:1967a}. As a result, the VHE spectrum measured here by H.E.S.S. provides constraints on the EBL density (see, e.g., \citealt{aharonian:2006:hess:ebl:nature}; hereafter Aha06) at near-infrared wavelengths. As 1ES\,0347-121 is one of the most distant objects, for which a VHE spectrum is measured, these EBL constraints are potentially very strong.
The broad-band SED of BL Lac objects, typically a double-peaked shape, can often be explained by SSC models (see, e.g., \citealt{aharonian:2005:hess:pks2155mwl}). However, the accuracy of such modeling is highly dependent on the observations being contemporaneous due to the extreme flux-variability of BL Lacs (e.g. \citealt{krawczynski:2000a}). The simultaneous UV/optical, X-ray, and VHE measurements presented here enable the first SSC modeling of the emission from this object including data in the higher-energy peak.

\section{H.E.S.S. observation and results}

\begin{figure}[tbh]
\centering
\includegraphics[width=0.45\textwidth]{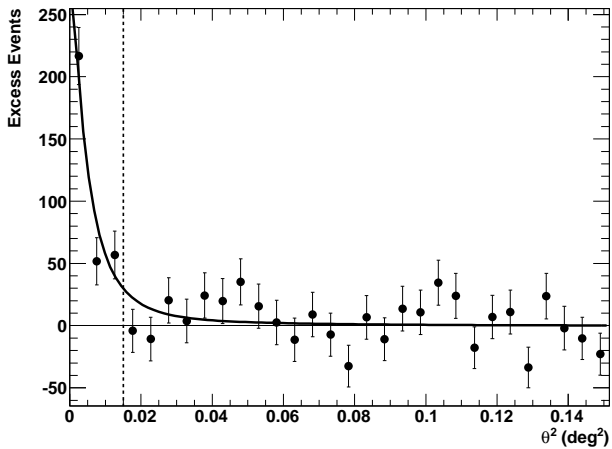}
\caption{The number of excess events versus squared angular distance $\theta^2$ to the nominal source position of 1ES\,0347-121. The solid line shows the expectation for a point source from Monte Carlo simulations with the same pointing offset and zenith angle distributions as the observations. The vertical dashed line shows the cut on the distance from the source used for signal extraction ($\theta^2 < 0.015$\,deg$^2$). }
\label{Fig:Theta2}
\end{figure}
 
\begin{figure}[tbh]
\centering
\includegraphics[width=0.5\textwidth]{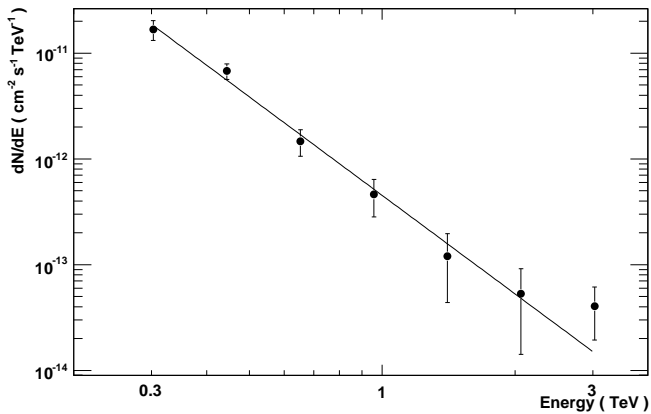}
\caption{The differential photon spectrum of 1ES\,0347-121. The line is the best fit of a power law to the data ($\Gamma =  3.10 \pm 0.23_{\mathrm{stat}} \pm 0.10_{\mathrm{sys}}$).}
\label{Fig:Spectrum}
\end{figure}

\begin{figure}[tbh]
\centering
\includegraphics[width=0.45\textwidth]{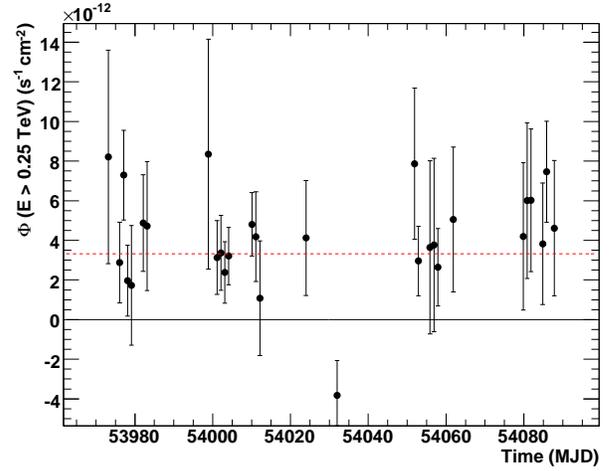}
\caption{The average nightly flux above 250 GeV from 1ES 0347-121. The dashed line shows the fit of a constant to the lightcurve ($\Phi(E > 250\mathrm{\,GeV}) = (3.32 \pm 0.44_{\mathrm{stat}} ) \times 10^{-12}$\,cm$^{-2}$\,s$^{-1}$; $\chi^2/\mathrm{d.o.f.} =  30.4/28 = 1.1$).}
\label{Fig:Lightcurve}
\end{figure}
  
The H.E.S.S. array of four imaging atmospheric-Cherenkov telescopes  \citep{hinton:2004a:hess:status} is used to search for VHE $\gamma$-ray emission from objects such as BL Lacs.
The H.E.S.S. observations of 1ES\,0347-121 were performed between August and December 2006. A total of 25.4\,h live time of good-quality data was recorded at zenith angles ranging from 12$^\circ$ to 40$^\circ$. The mean zenith angle is $\sim$19$^\circ$, for which the energy threshold of this analysis is $\sim$250\,GeV. These data are analyzed with a standard Hillas-type analysis \citep{aharonian:2006:hess:crab}. The data were recorded with a pointing offset of 0.5$^\circ$ relative to the nominal source position to allow a simultaneous estimation of the background using events from the same field of view (\textit{reflected background} from \citealt{aharonian:2006:hess:crab}).
 
An excess of 327\,$\gamma$-ray candidates is measured (1167 signal events, 9241 background events, background normalization: 0.0909) corresponding to a statistical significance of 10.1 standard deviations (following  Eq.~17 of \citealt{li:1983a}). Figure~\ref{Fig:Theta2} shows that the extension of the excess is compatible with that expected from a point-source. The fit position of the excess ($\alpha_{J2000}=3^{\mathrm h}49^{\mathrm m}23.0^{\mathrm s}\pm1.4^{\mathrm s}_{\rm stat}\pm1.3^{\mathrm s}_{\rm syst}$, $\delta_{J2000}=-11^{\circ}58'38''\pm33''_{\rm stat}\pm20''_{\rm syst}$) coincides with the location of 1ES\,0347-121 ($\alpha_{J2000}=3^{\mathrm h}49^{\mathrm m}23.2^{\mathrm s}$, 
$\delta_{J2000}=-11^{\circ}59'27.0''$;  \citealt{schachter:1993a}). The angular distance between the fit position of the VHE excess and the source position is $47''$.
The differential photon spectrum of the source is shown in Fig.~\ref{Fig:Spectrum}%
\footnote{An electronic version of the HESS spectrum is available from the publications section of the HESS website: http://www.mpi-hd.mpg.de/hfm/HESS/HESS.html.}%
. A fit of a power-law function $dN/dE = \Phi_0 (E/\mbox{1\,TeV})^{-\Gamma}$ to these data results in a statistically good description ($\chi^2/\mathrm{d.o.f.} = 3.5 / 5$) with normalization $\Phi_0 = (4.52 \pm 0.85_{\mathrm{stat}} \pm  0.90_{\mathrm{sys}}) \times 10^{-13}$\,cm$^{-2}$\,s$^{-1}$\,TeV$^{-1}$ and photon index $\Gamma = 3.10 \pm 0.23_{\mathrm{stat}} \pm 0.10_{\mathrm{sys}}$. The integral flux above 250\,GeV taken from the spectral fit is $\Phi(E > 250\mathrm{\,GeV}) = (3.9 \pm 1.1_{\mathrm{stat}} ) \times 10^{-12}$\,cm$^{-2}$\,s$^{-1}$, which corresponds to $\sim$2\% of the flux of the Crab Nebula above the same threshold \citep{aharonian:2006:hess:crab}.
As can be seen from Fig.~\ref{Fig:Lightcurve}, no significant variability is detected on time-scales of days or months. 

\section{SWIFT and ATOM observations and results}

\begin{figure*}[tbh]
\centering
\begin{minipage}[c]{0.65\textwidth}
\includegraphics[width=\textwidth]{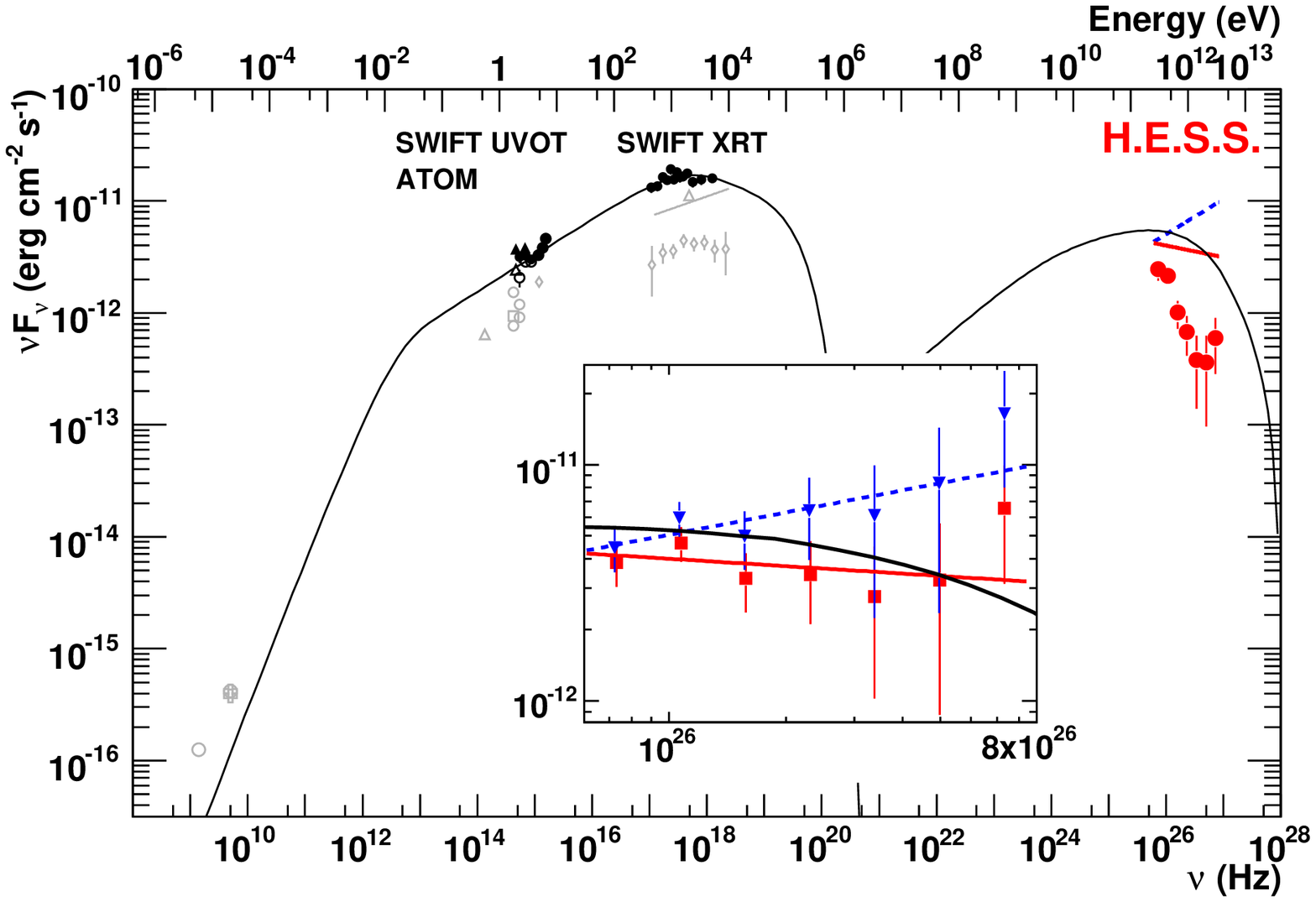}
\end{minipage}%
\begin{minipage}[c]{0.35\textwidth}
\caption{The spectral energy distribution of 1ES\,0347-121. Shown are the H.E.S.S measured spectrum (red markers), power-law fits to the intrinsic spectra corrected for the attenuation by different EBL models  (solid red line - \citealt{primack:2005a}; dashed blue line - P0.45 shape from \citealt{aharonian:2006:hess:ebl:nature}), and contemporary SWIFT XRT and UVOT (black circles) and ATOM measurements (black triangles) (open black markers: contribution from the host galaxy subtracted). Grey markers and lines are archival data \citep{elvis:1992a,schachter:1993a,perlman:1996a,wolter:1998a,raiteri:1998a,urry:2000a,cheung:2003a,giroletti:2004a,perlman:2005a}. A simple synchrotron self Compton model (see text) is also shown (solid curve). The model parameters are: 
$\delta = 25$, B = 0.035\,G and R = 3.2$\times$10$^{16}$\,cm. The electron distribution is described by a broken power law with K = 1.26$\times$10$^{4}$\,cm$^{-3}$, $\gamma_{\mathrm{min}}$ = 1$\times$10$^{3}$, $\gamma_{\mathrm{max}}$ = 3$\times$10$^{6}$, $\gamma_{\mathrm{break}}$ = 3.55$\times$10$^{5}$, n$_1$ = 2.35, and n$_2$ = 3.35.
The inlay shows a zoom on the intrinsic VHE spectra.
}
\label{Fig:SED}
\end{minipage}
\end{figure*}

SWIFT \citep{gehrels:2004a} observations of 1ES\,0347-121 were performed on October 3, 2006. A total of 3.2\,ksec of screened data in photon-counting mode are analyzed.
For the analysis of SWIFT XRT and UVOT data  the HEASOFT~6.2 package with Xspec11 and the latest calibration (XRT: 2007-03-30, UVOT: 2006-11-16) was used.
The XRT data has been reprocessed with the updated calibration tables including new bad pixel tables. In order to avoid pile-up effects in the energy spectrum, the core pixels in the image were not used and the extraction region was chosen to be an annulus (e.g. \citealt{giommi:2006a}) with an inner radius of 5 pixels and an outer radius of 30 pixels (1 pixel = $2.36''$). The background spectrum was accumulated in a wider annulus (inner radius of 45 pixels, outer radius of 90 pixels). The auxiliary response file was created with the standard tool \texttt{xrtmkarf} including the point-spread-function correction.
A power-law model fit to the data (response matrix \texttt{swxpc0to12\_20010101v008.rmf}; photoelectric absorption fixed to the Galactic value of $n_H=3.6\times 10^{20}$\,cm$^{-2}$) between 0.3 and 8\,keV yields a good $\chi^2/\mathrm{d.o.f.} = 105.3/105 = 1.00$ with $\Gamma=1.99\pm0.06$ (error at 90\% c.l.). The unabsorbed integrated energy flux between 2 and 10\,keV from the model is $f_{2-10\,\mathrm{keV}}= (2.800\pm0.003) \times 10^{-11}$\,erg\,cm$^{-2}$\,s$^{-1}$.

The UltraViolet/Optical Telescope (UVOT) observations were made using 6 filter settings. Since none of the light curves from the different measurements indicate variability, the data sets for the individual filter wheel settings were added together. The aperture was chosen to be of 12 pixels radius for the optical filters while it is increased to 24 pixels for the UV filters (1 pixel = $0.48''$). For the UV filter images, a 6-pixel radius region centered on a faint stellar source $12''$ north of 1ES\,0347-121 falling into the wider-24 pixel aperture was excluded. The aperture photometry includes a pile-up correction and the flux has been calculated from the count rates by using the zero point values quoted in the calibration notes of the SWIFT UVOT team. Finally, the observed flux was corrected for galactic absorption using a reddening of $E(B-V)=0.047$ which was then scaled to obtain $A_\lambda$ for the SWIFT filters following the recipe given by \citet{giommi:2006a}.

The Automatic Telescope for Optical Monitoring (ATOM; \citealt{hauser:2004a}) on the HESS site monitored 1ES\,0347-121 during the HESS observing period in November 2006.
The mean B-band flux measured is  $3.8 \times 10^{-12}$\,erg\,cm$^{-2}$\,s$^{-1}$ and the mean R-band flux is  $3.7 \times 10^{-12}$\,erg\,cm$^{-2}$\,s$^{-1}$ (aperture radius: $4''$). Both fluxes are constant, as all measurements are within 10\% and 4\%, respectively, of the mean values.

The resulting SWIFT X-ray and ultraviolet-to-optical (UV/O) and ATOM data are shown in Fig.~\ref{Fig:SED}.
The X-ray and UV/optical fluxes observed are the highest measured from this source today, indicating a period of enhanced activity during the H.E.S.S. measurements.

To correct for the contribution from the host galaxy in the UV/optical-bands  the method presented in \citet{aharonian:2007:hess:1es1101} is followed. The host galaxy flux in the R-band of $m_r = 17.26$ and the half-light radius $r_e = 1.25''$ are taken from \citet{urry:2000a}. Using a de Vaucouleurs profile, the flux from the host galaxy falling in the signal aperture is estimated to be $\sim$80\% for ATOM and $\sim$90\% for UVOT. The host galaxy flux in the V, B and U-band are estimated using the elliptical galaxy spectral template at $z = 0.2$ from \citet{fukugita:1995a}. For the UVOT data host galaxy contributions of $\sim$35\% (V-band), $\sim$11\% (B-band) and $\sim$4\% (U-band) are derived. For the ATOM data the host galaxy contributions are $\sim$34\% (R-band) and $\sim$8\% (B-band). The UV/optical measurements corrected for the contribution by the host galaxy are shown as open black markers in Fig.~\ref{Fig:SED}. 

\section{Discussion}

\begin{figure}[tbh]
\centering
\includegraphics[width=0.5\textwidth]{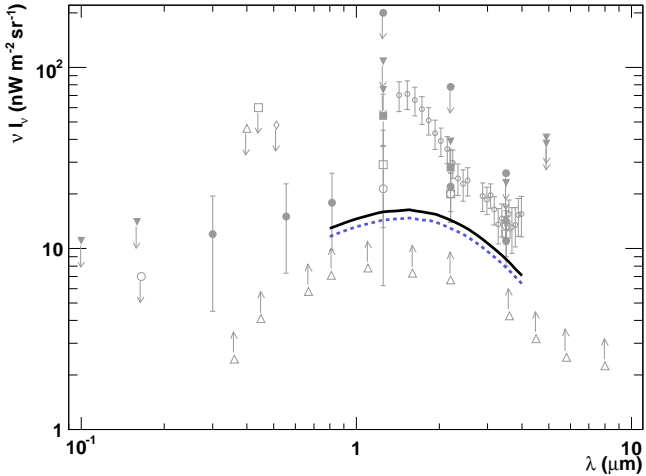}
\caption{The spectral energy distribution of the extragalactic background light (EBL). The black line is the limit (P0.61) on the EBL density derived from the measured spectrum of 1ES 0347-121 using the same technique as described in \citet{aharonian:2006:hess:ebl:nature}. The dashed curve is the limit derived in \citet{aharonian:2006:hess:ebl:nature} from the HESS spectrum of 1ES 1101-232. The data compilation (grey markers) is from \citet{mazin:2007a}.  }
\label{Fig:EBLLimit}
\end{figure}
 
The diffuse extragalactic photon field in the ultraviolet to far-infrared wavelength region (EBL) consists of the photons emitted by stars and absorbed and re-emitted by dust redshifted and integrated over time (see, e.g., \citealt{hauser:2001a} and \citealt{Kashlinsky2005:EBLReview} for recent reviews). Its spectrum carries cosmological information about galaxy and star formation history.
 
Distant sources of VHE $\gamma$-rays can probe the EBL density \citep{stecker:1992a}.  VHE photons passing through the EBL are attenuated via pair production: $\gamma_{\mathrm{VHE}} + \gamma_{\mathrm{EBL}} \rightarrow e^{+} + e^{-}$ \citep{gould:1967a}. Since this process is energy dependent, the VHE $\gamma$-ray spectra measured on Earth carry an attenuation imprint from the EBL. With reasonable assumptions about the emission physics of the source, limits on the EBL density can be derived.
 
Using the relatively hard energy spectra measured from the recently discovered VHE BL Lacs 1ES\,1101-232 (z = 0.186) and H\,2356-309 (z = 0.165), strong constraints on the EBL density in the optical to near-infrared ($\sim$0.8-4\,$\mu$m) were derived (Aha06). Following the exact same methodology as described in Aha06 (scaling of a reference EBL shape until the intrinsic spectrum reaches a maximum hardness) a limit on the EBL density is derived from the observed VHE spectrum of 1ES\,0347-121. For a scaling factor of 0.51 (i.e. a P0.51 shape), the fit of a power-law function to the intrinsic spectrum results in a photon index of $\Gamma = 1.5$, the minimum allowed value. Taking into account evolution of the EBL, again using the same arguments as Aha06, a limiting shape of P0.61 is derived. The limit is shown in Fig.~\ref{Fig:EBLLimit} in comparison to the limit derived by Aha06 for 1ES\,1101-232 (P0.55). The limit derived here is only slightly less constraining, which is a result of the softer observed spectrum and the smaller energy range towards low energies. The claimed excess of the EBL in the near-infrared above the value derived from source counts \citep{matsumoto:2005a}, often attributed to a possible contribution of the first stars to the EBL, is again excluded.

Given the large distance of the source, the measured VHE spectrum is severely altered by the EBL attenuation. Therefore, the VHE spectrum has to be corrected for this attenuation prior to any modeling of the intrinsic SED. Since the exact shape of the EBL is not known, the intrinsic spectrum is calculated for two different EBL models: the model from \citet{primack:2005a} (PRIM), which has a low EBL density, and the upper limit shape P0.45 from Aha06 (AH). The intrinsic VHE spectra (inlay in Fig.~\ref{Fig:SED}) are each well fit by power-law functions, which are shown in Fig.~\ref{Fig:SED} above the measured VHE photon spectrum. The photon indices of the power-law functions are $\Gamma = 2.10 \pm 0.21$ for the PRIM and $\Gamma = 1.69 \pm 0.22$ for the AH shape.

A simple homogeneous one-zone synchrotron self Compton (SSC) model from
\citet{krawczyski:2004a}
is used to describe the SED from the contemporaneous UVOT, X-ray and VHE data. For the modeling, the data are not strongly constraining. Parameter sets describing the overall spectral shape with a standard shock-accelerated particle distribution can be found with a good statistical compatibility between the host-galaxy-substracted UVOT data, the X-ray data, both intrinsic VHE spectra and the model. In Fig.~\ref{Fig:SED} an SSC model, fit to the host-galaxy-substracted UVOT, X-ray, and the intrinsic VHE data resulting from the AH EBL shape, is shown ($\chi_{\mathrm{red.}}^2 \sim 1.2$; $P(\chi^2) \sim 0.25$).
The archival radio measurements are assumed to be produced by a different particle population and are not included in the fit.
The model parameters, although not strongly constrained, are similar to the parameters used previously to model other BL Lacs (e.g. \citealt{aharonian:2007:hess:1es1101}).

To summarize, the new distant TeV blazar 1ES\,0347-121 is detected at energies $>$250\,GeV with the H.E.S.S. experiment. It is among the most distant TeV blazars discovered today. The relatively hard VHE spectrum confirms previous limits on the EBL density at near-infrared wavelengths. X-ray and UV/O data, taken during the H.E.S.S. observation period with the SWIFT satellite, indicate a high flux state of the source. The contemporaneous X-ray, UV/O, and VHE data are not strongly constraining and can be described by a simple SSC model. Further multi-wavelength observations, especially in X-rays, can improve the constraints on the model parameters. Many TeV blazars show large variations in the VHE flux, so monitoring the source, to measure a state of high VHE flux (flare), is desirable.


\begin{acknowledgements}
The support of the Namibian authorities and of the University of Namibia
in facilitating the construction and operation of H.E.S.S. is gratefully
acknowledged, as is the support by the German Ministry for Education and
Research (BMBF), the Max Planck Society, the French Ministry for Research,
the CNRS-IN2P3 and the Astroparticle Interdisciplinary Programme of the
CNRS, the U.K. Science and Technology Facilities Council (STFC),
the IPNP of the Charles University, the Polish Ministry of Science and 
Higher Education, the South African Department of
Science and Technology and National Research Foundation, and by the
University of Namibia. We appreciate the excellent work of the technical
support staff in Berlin, Durham, Hamburg, Heidelberg, Palaiseau, Paris,
Saclay, and in Namibia in the construction and operation of the
equipment.
The authors wish to thank G. Tagliaferri, N. Gehrels and the SWIFT team for the support and cooperation in triggering the SWIFT observations.
The authors also wish to thank the referee for helpful comments and suggestions.
This research has made use of data obtained through the High Energy Astrophysics Science Archive Research Center Online Service, provided by the NASA/Goddard Space Flight Center,
and of NASA's Astrophysics Data System.
\end{acknowledgements}


\bibliographystyle{aa}

\bibliography{1es0347_hess_paper_v0.6.3}
 
\end{document}